\documentclass[epj]{webofc}
\usepackage[varg]{txfonts}   
\woctitle{MENU 2013}
\begin{document}
\title{SPQR ---  Spectroscopy: Prospects, Questions \& Results}
%
%

\author{M.~R.~Pennington\inst{1}\fnsep\thanks{\email{michaelp@jlab.org}}
}

\institute{Theory Center, Thomas Jefferson National Accelerator Facility, 12000 Jefferson Avenue, Newport News, VA 23606, U.S.A. 
}

\abstract{%
Tremendous progress has been made in mapping out the spectrum of hadrons over the past decade with plans to make further advances in the decade ahead. Baryons and mesons, both expected and unexpected, have been found, the results of precision experiments often with polarized beams, polarized targets and sometimes polarization of the final states. All these hadrons generate poles in the complex energy plane that are consequences of the strong coupling regime of QCD. They reveal how this works.
}
\maketitle
\section{Why Spectroscopy?}
\label{intro}
The spectrum of states of any system is fundamental: reflecting the constituents that make up that system and the interactions between them. The rich spectrum of hadrons reveals the workings of QCD in the strong coupling regime. There are two ways to study this. One is wholly theoretical. Knowing the QCD Lagrangian as we do, one can, in principle, compute its consequences. This turns out to be only just within our capabilities, and only in simpler cases can definitive results be obtained. The alternative is to use experiment as our guide, and learn from there. In experiment quarks know how to solve the field equations of QCD in the strong coupling regime even without the help of  a BlueGene computer. Nevertheless, extracting the spectrum from complex data is often far from straightforward, requiring close interaction between theory and experiment.

Substantial progress has been made in both the baryon and meson sectors during the past ten years with increasingly precise experiments, measuring not just differential cross-sections, but all manner of polarization observables too. Even more results are to come from BESIII, COMPASS, LHCb,
 MAMI, ELSA, and Jefferson Lab experiments, with PANDA to follow. 

\section{The Hadron Spectrum: Baryons and Mesons}
\label{sec-1}


\begin{figure}[t]
\centering
\includegraphics[width=0.8\textwidth]{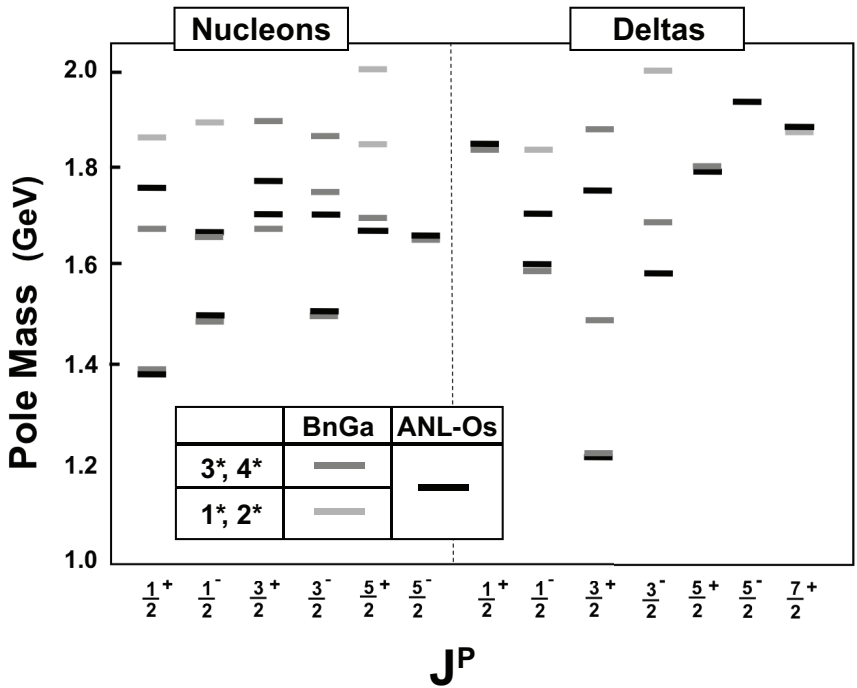}
\vspace{-2mm}
  \caption{$N^*$ and $\Delta^*$ spectra, labeled by their spin and parity as $J^P$ along the abscissa, and the real part of the resonance pole positions along the ordinate, from the Bonn-Gatchina~\cite{bn-ga} and  ANL-Osaka~\cite{ebac2} analyses of experimental data. For the ANL-Osaka ({\it aka} EBAC) analysis all the states have $3^*-4^*$ provenance, while Bonn-Gatchina also include those with $1^*-2^*$ ratings, according to the legend shown.
Note the tendency of some  $N^*$'s and $\Delta^*$'s to appear in parity pairs as their mass increases above 1800 MeV~\cite{bn-ga-doublet}.}
\label{fig-1}
\vspace{-7.5mm}
\end{figure}

Baryons have a special place in the firmament of quark bound states. First it was their multiplet structure that led to the proposal of the quark model, and the discovery of the triply strange $\Omega^-$ that confirmed this was on the right lines. The inclusion of quarks in the dynamics of QCD made baryons special too.  They most obviously reflect the non-Abelian nature of the theory, since a minimum of three quarks each with different colour charges is required to build a colour neutral hadron with half-integer spin. To learn about the spectrum of excited baryons we first fired pion beams at proton targets and measured the cross-section and polarization for the production of $\pi N$ and $\pi\pi N$ final states. Since states in the spectrum of hadrons have definite quantum numbers, to find these the $\pi N$ cross-sections and asymmetries are decomposed into underlying amplitudes with definite spin.
However, these only provided a glimpse of a limited part of the spectrum. A more complete picture is provided by detecting strange, as well as non-strange, final states (like $K\Lambda$, $K\Sigma$, {\it etc})~\cite{sarantsev,bn-ga2,menu} and  by more recent studies with photon beams, in different polarization states scattering on polarized targets~\cite{bn-ga2}. This has been enabled by a wonderful set of experiments at ELSA@Bonn, MAMI@Mainz and CLAS6@JLab. The outcome of two Amplitude Analyses of all these data is shown in Fig.~1. One is a sophisticated, but traditional Amplitude Analysis, by the Bonn-Gatchina team~\cite{bn-ga}, and the other which attempts to learn about the underlying dynamics directly is that by the ANL-Osaka group~\cite{ebac2}. This uses the Sato-Lee effective Lagrangian~\cite{sato-lee} as its basis, and relies on computing the contribution of many Feynman diagrams as the energy increases. While these approaches satisfy unitarity for two-body channels, three- and higher-body interactions are more complicated. Consequently, it is the more flexible Bonn-Gatchina analysis that can fit the $\pi\pi N$ final states and determine the spectrum to higher masses. The results in Fig.~1 show that the $N^*$'s and $\Delta^*$'s from these two analyses have much in common, but there are some key differences that need to be resolved. The measurement of double polarization asymmetries, like the so called $G$-function with linearly polarized photons on a longitudinally polarized target open a unique window on to the higher partial waves~\cite{krusche,bn-ga3}. They show that the need for important spin-3 components above 1.55~GeV, seen in the top right corner of Fig.~1. Many of these new results from Bonn and Mainz are being presented at this conference~\cite{krusche,thiel}. More data are to come. Beam and target technology are providing detailed access to this spectrum up to 2.2~GeV.

The aim is not just to assemble hadron states like a stamp  collection, but to determine their masses and widths (given by their poles in the complex energy plane), and their couplings to all the channels in which they appear (given by the appropriate residues of these poles), and from these to learn about the composition of these states.
By virtue of the uncertainty principle, the proton and neutron inevitably have a meson cloud, which has detectable effects --- much like the Lamb shift in QED. However, for excited states this cloud is even more tangible. It is real. $\pi N$ and $\pi\pi N$ configurations are an essential part of the Fock space of the $N^*(1440)$ and all the many excited states shown in Fig.~1. It is through these components that each decays. The degrees of freedom are not just three quarks, but all the decay channels too. They are not just objects with a $qqq$-core of the constituent quark model~\cite{capstick}, but they must have additional ${\overline q}q$, or even ${\overline q}gq$ components. The aim is to determine this structure for each of the lower lying excited states, and then to understand from this the detailed workings of strong coupling QCD.
Studying in electroproduction experiments  how these compositions change as the virtuality of the probing photon increases, may yet confirm these insights~\cite{mokeev}. 

In the constituent quark model, decays were often treated as some \lq\lq perturbative'' addition, as in the $^3P_0$ scheme~\cite{barnes}. However, more recently, it has been appreciated that decays 
actually change the dynamics of the spectrum~\cite{pennington-wilson, santopinto}. This complexity can bring new states into view, for which the opening of decay channels are essential, while making others merge into the continuum as they no longer bind but just fall apart. Such hadronic components are there in modern lattice calculations too~\cite{edwards1}. However at present with $up$ and $down$ quarks having 10 times their physical mass, and so pions of $400$~MeV, only to a limited extent. As computations advance towards pions of 140~MeV, these hadronic components are likely to shift the masses of the resulting baryons and change their couplings~\cite{mrp-LEAP}, hopefully, approaching those that appear in experiment. 


That decay channels are essential to hadron states has long been suspected for mesons: the enigmatic scalars~\cite{mrp-FSU} $f_0(980)$ and $a_0(980)$ clearly have ${\overline K}K$ channels at the heart of their existence. The discovery of the new $X,\,Y,\,Z$ states in the heavy quark sectors have highlighted this too. The $X(3872)$ is closely associated to the $D {\overline D}^{*}$ channel. The charged $Z_c(4430)$ clearly must be more complex than simply ${\overline c}c$. New states with hidden strangeness have been found too, like the $Y(2175)$ in the $\phi f_0(980)$ channel. These all have the feature that $S$-wave coupling to nearby hadron channels brings binding.

Indeed, it is in the meson sector where some of the previously unconfirmed QCD configurations of colour singlets are to be found: glueballs and hybrids. A world of pure glue, while theoretically most interesting, doesn't exist in the real world. Light glueball configurations inevitably mix with channels in which ${\overline q}q$ states appear through their common $\pi\pi$, ${\overline K}K$, $\eta\eta$, etc., decay channels. However, hybrids, states in which glue contributes not just binding but to their quantum numbers, can arise with $J^{PC}$'s not possible for simpler ${\overline q}q$ systems. Such states like $1^{-+}$ are called \lq\lq exotic'', but they are only exotic in the quark model, not in QCD, where their appearance is to be expected.


\begin{figure}[t]
\centering
\includegraphics[width=0.90\textwidth]{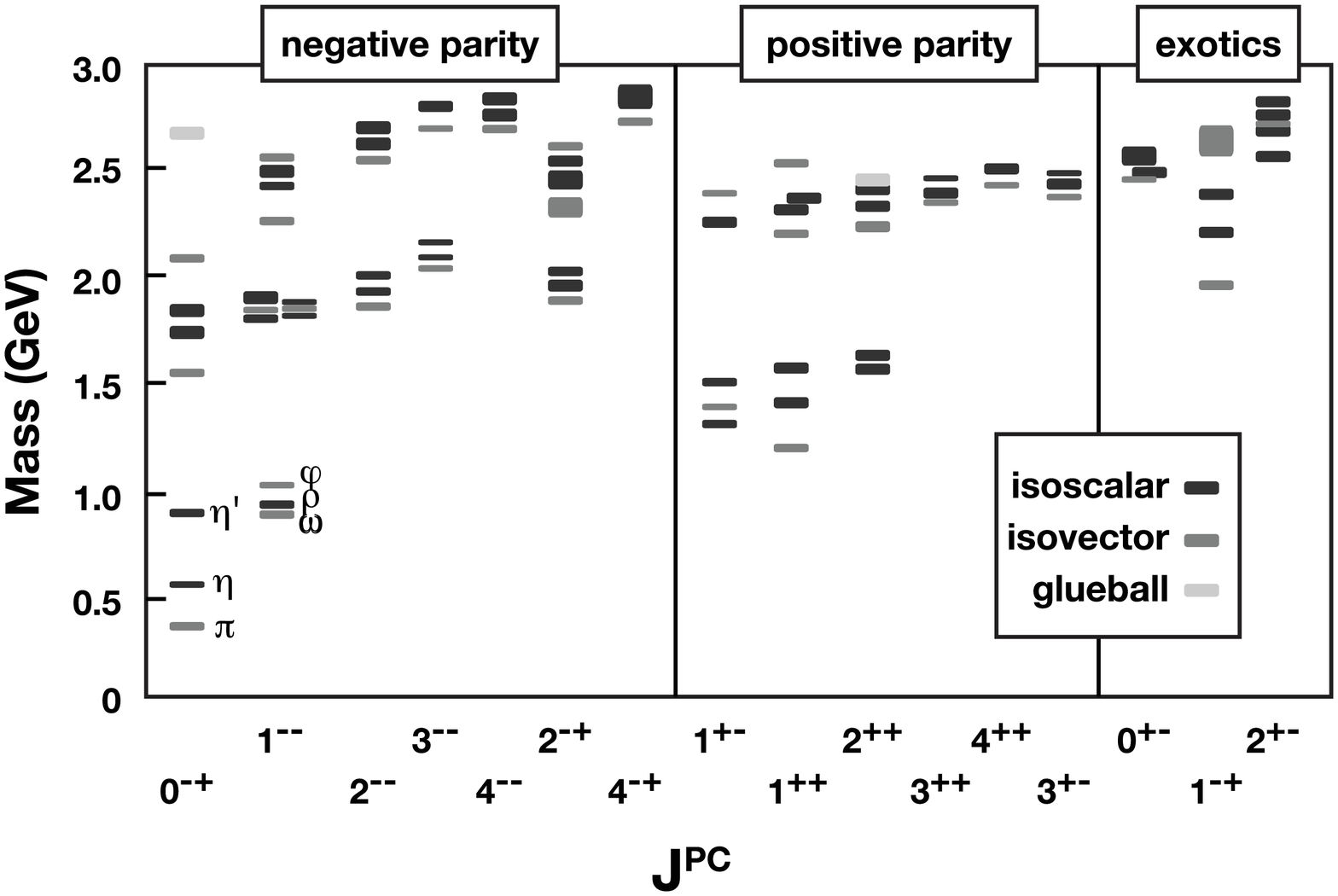}
  \caption{Lattice QCD results for the meson spectrum labeled by their spin and parity as $J^{PC}$ along the abscissa, and their masses along the ordinate, from the Hadron Spectrum Collaboration~\cite{dudek} with a pion mass of 400~MeV, showing their flavour structure. The calculations are for states constructed from  operators with $q$, $\overline{q}$ and $g$
configurations. The results are grouped into those with natural and unnatural parity allowed by simple ${\overline q}q$ states. Those labeled {\it exotic} do not appear in the quark model, and in the lattice calculation are dominated by ${\overline q}gq$ components.}
\label{fig-2}
\vspace{-4.5mm}
\end{figure}

The latest lattice calculations~\cite{dudek}, shown in Fig.~2, predict multiplets of such states around 2~GeV. Since these computations are in a world with 400~MeV pions, they are expected to be shifted in the real world, just as we discussed for baryons. Nevertheless, the calculations are robust enough for a whole new program of exploration to be the focus of the Hall D program at Jefferson Lab~\cite{whitepaper}. There polarized photons scattering on a nucleon target will be studied in many final states: $\pi \pi N$, $3\pi N$, $\eta N$, $4\pi N$, $5 \pi N$, $\eta' N$, etc, with a detector designed to have a close to perfect acceptance. To this will be added kaon identification. With millions of events, the aim is to perform precision partial wave analyses. Hybrids, and other new states involving light flavours of quark, are unlikely to be narrow, and appear as simple \lq\lq bumps'', but only by performing Amplitude Analyses of many channels simultaneously will poles in the complex energy plane be definitively revealed. This requires close cooperation between theorists and experimentalists. To facilitate this, the  JLab Physics Analysis Center has been set up, led by Adam Szczepaniak. 
\section{JLab Physics Analysis Center}
\label{sec-6}
The states that first populated the Particle Data Tables were those that naturally were those that lived longest and so appearing as narrow(ish) peaks in the appropriate integrated cross-section: the $\rho$, $\omega$, $\phi$, $N^*(1520)$, $\cdots$. This gave the impression that determining the hadron spectrum was just a matter of bump-hunting. However, it soon became clear that many states were highly inelastic, appearing in several channels, often not creating more than a wiggle in any one cross-section. Nevertheless, these correspond to poles in the complex energy plane, which is the true signature of a state in the spectrum of states. Others, like the $f_0(980)$, couple strongly to a threshold that is just about to open above their notional mass. Such a state appears as a peak in some reactions and as a dip in others. Nevertheless, these too are described by a pole in the complex energy plane, regardless of the way they appear in experiment on the real energy axis. All this makes it clear that one must have the right framework in which to describe the amplitudes in which resonances appear and the right tools to continue the amplitudes into the complex energy plane. This framework is provided by Reaction Theory. This requires that the Scattering (or $S$-) matrix that describes each reaction satisfies the consequences of causality, relativity and the conservation of probability. These are the basics of no particular theory, but every theory. These require that the $S$-matrix elements possess the correct  analyticity, crossing and unitarity properties.

Amplitudes are complex functions. Experiment can sometimes determine both their modulus and phase, or at least their relative phase. To connect these from one energy to another demands the use of
dispersion relations, or other analytic mapping techniques. Our experimental colleagues, who conceive and build the detectors and understand their acceptances, write the data acquisition software, connect up the electronics and a thousand myriad things to turn pulses into cross-sections, need the help of theorists to provide the translation of these results into the physics of hadrons. Theorists are an integral part of the analysis team, increasingly embedded within the collaborations. The aim here is not to prove some particular favourite model, whether based on constituent quarks, or some modelling of interactions in the bound state equations, or even to validate a lattice calculation, but rather to input essential truths of scattering theory. Testing models has a role, but that comes later, once definitive results have been obtained from experiment.


\begin{figure}[h]
\vspace{1.5mm}
\centering
\includegraphics[width=0.70\textwidth]{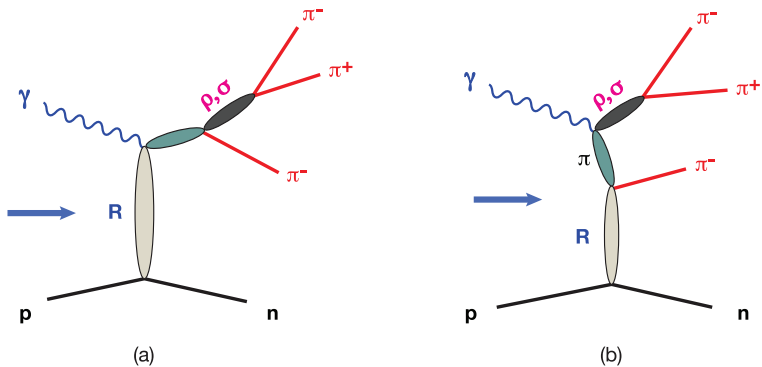}
  \caption{Two of the reaction mechanisms at work in $\gamma N \to 3\pi N$. (a) represents Regge exchange (R) creating intermediate states that decay to $\,\rho\pi$ or $\sigma\pi$, that might include $1^{-+}$ quantum numbers. (b) Deck production of the same final state. These mechanisms interfere. The consequences of this have to be understood across the kinematic range of the reaction to determine the production mode of any partial wave.}
\label{fig-3}
\vspace{-4.5mm}
\end{figure}

The mission of the JLab Physics Analysis Center is to network with appropriate theorists and experimentalists in different collaborations to achieve this goal, whether with CLAS12 or GlueX@JLab, COMPASS@CERN, BESIII@BEPC or PANDA@Fair. The purpose of this networking for spectroscopy is to share the $S$-matrix technology that is required  and to make this a practical tool. To this end, various working groups have been set up for the first year to study reaction mechanisms and final state interactions, in particular. As prompted by the discussions of the $a_1$ years ago, multi-hadron production is far from simple. To establish that the $a_1$ was indeed a state in the spectrum  required a detailed understanding of how the different mechanisms for three pion production contributed, Fig.~3; whether the behaviour of the relevant $J^{PC}\,=\,1^{++}$ $3\pi$ partial wave requires a resonance  like that generated by the graph in Fig.~3a, or can it be wholly understood in terms of the Deck effect of Fig.~3b. Multi-body final state interactions play a key role in searching for new states that may point to glue as an essential contributor to their $J^{PC}$ quantum numbers. Heavy flavour factories, like BaBar and Belle are rich sources of information about such decays. This has to be combined with practical methods for implementing two and three-body unitarity to be used in Amplitude Analyses of the precision data to come. COMPASS is confronting all these issues~\cite{compass1} and is a key experiment from which we hope to learn. To meet these demands the JLab Physics Analysis Center is not just working with experimentalists but establishing close connections with other theory consortia like the NABIS group~\cite{nabis} and the Haspect project~\cite{haspect}. To make the most of the precision data that modern experiments deliver, with much more to come, we must have tools of comparable precision to extract the detailed physics required to understand how the dynamics of QCD, with its properties of colour confinement and chiral symmetry breaking, really works.  That is the challenge.

\newpage
%
It is a pleasure to thank the organisers, especially Annalisa D'Angelo,
 for the invitation to give this talk in such an auspicious venue. Discussions with Reinhard Beck on the latest experimental results were much appreciated.
This paper has been authored by Jefferson Science Associates, LLC under U.S. DOE Contract No. DE-AC05-06OR23177.

%
%
%

%
%

\end{document}